\documentclass[prl,twocolumn,superscriptaddress,floatfix,nobalancelastpage]{revtex4}%
\usepackage{graphicx,bm}%
\usepackage{amsmath}%
\usepackage{amsfonts}%
\usepackage{amssymb}%
\usepackage{graphicx}

\providecommand{\U}[1]{\protect\rule{.1in}{.1in}}

\begin{document}

\title{Reply to Comment by Borisenko \emph{et al.} on article `A de Haas-van Alphen
study of the Fermi surfaces of superconducting LiFeP and LiFeAs'}
\author{C. Putzke}
\affiliation{H.H. Wills Physics Laboratory, University of Bristol, Tyndall Avenue, Bristol, BS8 1TL, UK.}
\author{A.I. Coldea}
\affiliation{Clarendon Laboratory, Department of Physics, University of Oxford, Parks Road, Oxford OX1 3PU, U.K.}
\author{I. Guillam\'{o}n}
\affiliation{H.H. Wills Physics Laboratory, University of Bristol, Tyndall Avenue, Bristol, BS8 1TL, UK.}
\author{D. Vignolles}
\affiliation{Laboratoire National des Champs Magn\'{e}tiques Intenses (CNRS), Toulouse, France.}
\author{A. McCollam}
 \affiliation{High Field Magnet Laboratory, IMM, Radboud University Nijmegen, 6525 ED Nijmegen, The Netherlands.}
\author{D. LeBoeuf}
\affiliation{Laboratoire National des Champs Magn\'{e}tiques Intenses (CNRS), Toulouse, France.}
\author{M.D. Watson}
\affiliation{Clarendon Laboratory, Department of Physics, University of Oxford, Parks Road, Oxford OX1 3PU, U.K.}
\author{I.I. Mazin}
\affiliation{Code 6393, Naval Research Laboratory, Washington, DC 20375, USA.}
\author{S. Kasahara}
\affiliation{Research Center for Low Temperature and Materials Sciences, Kyoto University, Sakyo-ku, Kyoto 606-8501,
Japan.}
\author{T. Terashima}
\affiliation{Research Center for Low Temperature and Materials Sciences, Kyoto University, Sakyo-ku, Kyoto 606-8501,
Japan.}
\author{T. Shibauchi}
\affiliation{Department of Physics, Kyoto University, Sakyo-ku, Kyoto 606-8502, Japan.}
\author{Y. Matsuda}
\affiliation{Department of Physics, Kyoto University, Sakyo-ku, Kyoto 606-8502, Japan.}
\author{A. Carrington}
\affiliation{H.H. Wills Physics Laboratory, University of Bristol, Tyndall Avenue, Bristol, BS8 1TL, UK.}

\maketitle

Recently, Borisenko \textit{et al.} \cite{borisenkocomment} have posted a Comment where they suggest an alternative
interpretation of our de Haas-van Alphen (dHvA) measurements on the superconductor LiFeAs \cite{putzke11}. In our
original paper \cite{putzke11} we concluded that our measurements of the bulk Fermi surface were not consistent with
the surface bands observed thus far by ARPES. Borisenko \textit{et al.} \cite{borisenkocomment} dispute this and
suggest the two measurements are consistent if some of the orbits we observe are due to magnetic breakdown. We argue
here that this scenario is inconsistent with the experimental data and therefore that our original conclusion stands.

The Comment by Borisenko \textit{et al.}, \cite{borisenkocomment} consists of several claims. First, it is claimed that
the technique of interpreting the dHvA spectra in terms of the calculated band structure, and accounting for the
deviation from the calculated density functional (DFT) bands by assigning small mutual shifts to the bands, fails in
case of LiFeAs. To prove that, they plot together the \textit{unshifted} bands (our Fig. 2c-left) overlapping them with
the measured data of Fig. 2c-right. In fact, our Fig. 2c-right shows that the \textit{shifted} bands agree with the
experiment within the experimental noise. We note that the required shifts are surprisingly small, smaller than in any
other multiorbital metal, correlated or not, e.g., MgB$_{2}$ or Sr$_{2}$RuO$_{4}$ \cite{carringtonmgb2,mackenzie}. We
also note that, contrary to the claim in the Comment \cite{borisenkocomment}, although in general the hole and the
electron bands are allowed to shift independently in the experimental fitting procedure, the final result is checked
for physical consistency by ensuring the total number of electrons is conserved. In fact, the very small size of the
shifts needed in the case of LiFeAs renders this issue irrelevant.

Their second claim is that in addition to the aforementioned shifts a huge mass renormalization is needed to reproduce
the observed cross-section areas, depicted in Fig. 2c. This is a misconception. Effective masses have no effect on the
areas, and no other modifications of the calculated bands, besides small shifts, had been made in making Fig. 2c-right.
The authors of the Comment \cite{borisenkocomment} were perhaps confused by the fact that dHvA allows for a separate
extraction of the effective masses (from the temperature dependence of the peak amplitudes), and these appear to be
very orbital dependent, with one particular band showing mass renormalization of 4.9.

The authors of the Comment appear surprised that LiFeP (though not LiFeAs) requires up to 73\thinspace meV shifts to
describe the experimental results. However, this is in fact a very typical number for any multiband material. We recall
that MgB$_{2}$, which is exceedingly well studied by dHvA, ARPES, and many other techniques, and lacks magnetism,
correlations, and other complications of ferropnictides, requires shifts of up to 100\thinspace meV. On the other hand,
the parallel between LiFeAs and LiFeP, that these authors try to exploit here, is indeed extremely useful and
important. The bulk sample quality of LiFeAs as of now is not quite as good as that of LiFeP, but the fact that in
LiFeP all orbits below 3\thinspace kT are seen \textit{and} are reproduced exceedingly well by our procedure, leaves no
room to doubt the applicability of the same procedure for LiFeAs. The same procedure has also been used to fit in very
high detail other ferropnictides such as LaFePO \cite{amalia08}, SrFe$_{2}$P$_{2}$ \cite{analytis09}, BaFe$_{2}$P$_{2}$
\cite{arnold11} and BaFe$_{2}$As$_{2}$ \cite{Terashima11}. We note in passing that the authors appear to have another
misconception, that a band with a large mass renormalization would require a larger shift to achieve the same
adjustment in the orbit areas than an unrenormalized band. In fact the opposite is true; while we had to shift the DFT
band 1 by 60 meV, the renormalized band ($m^{\ast}/m=2.3$ for this band) would have to be shifted by 26 meV.

Yet another point of the Comment is that the observed near degeneracy of the 2a and 5a orbits is accidental. This is
indeed true, and we say that explicitly in our paper \cite{putzke11}. Moreover, DFT being only an approximation, the
real bands may or may not hold this degeneracy. However, the reader should be aware that the LMTO-ASA method to which
Borisenko \textit{et al.} \cite{borisenkocomment} refer is only a less accurate method for solving the same DFT
equations than the LAPW method used by us. In other words, a high accuracy solution of DFT equations yields this
accidental degeneracy, and a less exact solution of the same equations violates it. Of course, a more accurate
approximation than DFT may also violate it, but one cannot address this issue in the way the comment
\cite{borisenkocomment} does. We also note here that the authors of Ref.\ \cite{borisenkocomment} do not state if the
structural parameters used in the unpublished LMTO calculation they cite are the same as in ours. This could be another
source of discrepancy because the band-structure depends strongly on the As internal position.

\begin{figure}[ptb]
\center
\includegraphics*[width=8cm]{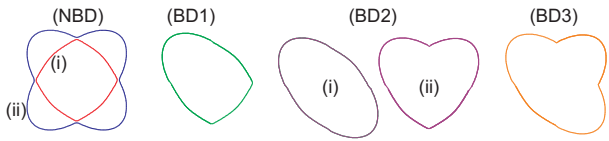}\caption{Possible orbits in the
electron bands of LiFeAs at $k_{z}=0.5c^{*}$. NBD shows the two non-breakdown orbits. The orbits resemble a flower with 4 petals.
The breakdown orbit BD1 has the central part plus 1 petal, BD2 has 2 additional petals and BD3 has 3 petals.}
\label{orbits}%
\end{figure}

The main substance of Ref.\ \cite{borisenkocomment} is the suggestion that the dHvA data might be consistent with the
ARPES results if magnetic breakdown is considered.   In the Comment, the authors compare our data to their unpublished
ARPES measurements where they have observed both electron Fermi surfaces as a function of $k_{z}$ (in Ref.\
\cite{borisenkoprl} only one electron surface and no $k_{z}$ dispersion was reported). In fact, in the comparison shown
in Fig. 1 of their Comment the authors use only the experimental Fermi surface cross-sections at the center and top of
the zone and then estimate the angle dependence of the dHvA frequencies using the result of the LMTO calculations
\textit{at low angle only} \cite{borispersonal}, extrapolating the low angle result to higher angle instead of using
the actual calculations. Such a procedure can hardly be considered to be rigorous but nevertheless they suggest that
the $\varepsilon$ dHvA orbit (see Fig.\ \ref{rotplot}) is consistent with a magnetic breakdown orbit coming from the
electron orbits at top of the zone and that the $\gamma$ and $\beta$ orbits come from the conventional (non-magnetic
breakdown) orbits from the inner electron sheet (at the top and center of the zone respectively). It is suggested that
this occurs because the gap between the electron sheets (which is induced by the spin-orbit interaction) is very small.
We had  carefully considered this scenario and concluded that it is inconsistent with the data.

\begin{figure}[ptb]
\center
\includegraphics*[width=8cm]{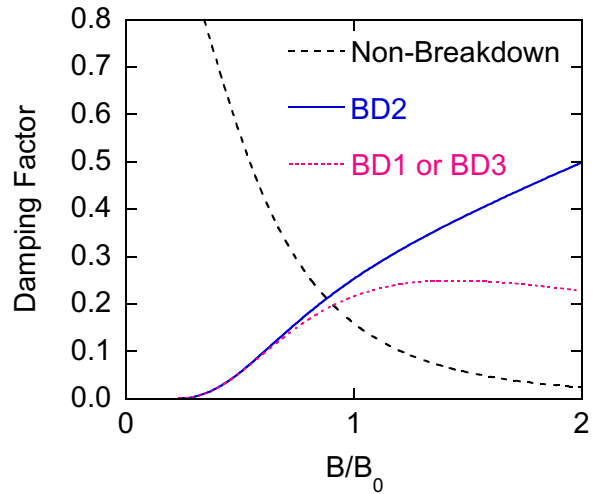}\caption{Relative probabilities of
performing the orbits shown in Fig.\ \ref{orbits}. Non breakdown = $(1-p)^4$, BD2 = $2p^4+4p^2(1-p)^2$, BD1 or BD3 = $4p^2(1-p)^2$, where $p=\exp(-B_0/B)$.}%
\label{probs}%
\end{figure}

Magnetic breakdown occurs when the electron has sufficient energy to tunnel between bands. The probability of it
occurring depends exponentially on the field, $p=\exp(-B_{0}/B)$, where the breakdown field $B_{0}$ depends on the size
of the gap. For the electron orbits in LiFeAs (see Fig.\ \ref{orbits}), for $B\ll B_{0}$ two orbits should be seen
corresponding to the inner and outer electron bands, whereas for $B\gg B_{0}$ a single orbit corresponding to the
ellipse in the unfolded zone is expected (BD2(i) in Fig.\ \ref{orbits}). In the intermediate field limit $B\simeq
B_{0}$ both types of orbit would be seen along with two further orbits corresponding to the inner orbit plus one or two
or three additional sections of the outer orbit (see Fig.\ \ref{orbits}). Because the ellipse orbit requires 4
breakdown gaps to be crossed, its probability is proportional to $p^{4}$, whereas the low field orbits are proportional
to $(1-p)^{4}$ and the intermediate orbits, which require two tunneling events and two reflections, are proportional to
$p^{2}(1-p)^{2}$. These probabilities, multiplied by their frequency of occurrence, are plotted as a function of
$B/B_{0}$ in Fig.\ \ref{probs}.

\begin{figure}[ptb]
\center
\includegraphics*[width=8cm]{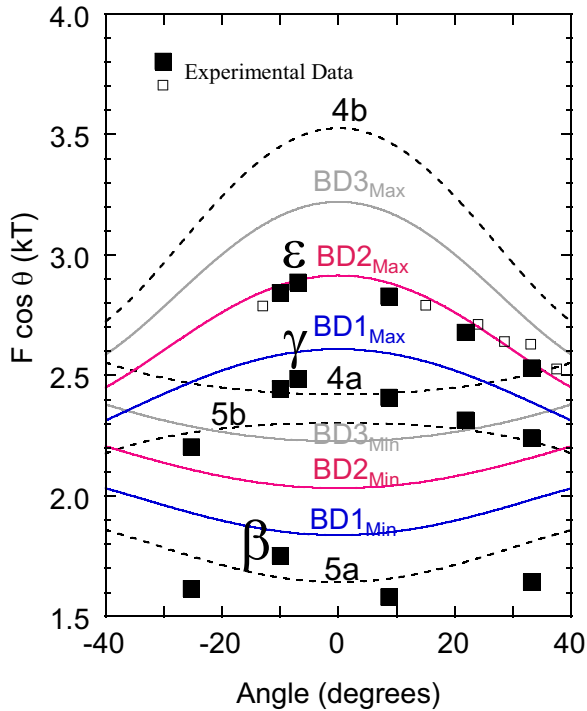}\caption{Experimental dHvA data along
with the ARPES bands (calculated in Ref.\ \cite{borisenkocomment}), showing
the non-breakdown orbits (dash lines) labelled as in Ref.\ \cite{putzke11} and
the various predicted breakdown orbits.  The greek letters, $\varepsilon$, $\gamma$ and $\beta$ label the three experimentally observed frequency branches.}%
\label{rotplot}%
\end{figure}

The interpretation suggested in Ref.\ \cite{borisenkocomment} requires the observation of both breakdown and
non-breakdown orbits so as shown in Fig.\ \ref{probs} we require $B\simeq B_{0}$, and hence all three types of
breakdown orbit should be observed with approximately equal amplitude. In additional, breakdown should be equally
likely to occur at the center and the top of the zone and hence the number of breakdown orbits is doubled. Using the
estimated ARPES $F(\theta)$ curves from Ref.\ \cite{borisenkocomment} we plot all these predicted orbits along with the
data in Fig.\ \ref{rotplot}. It can be seen that only the breakdown curve $\rm{BD2_{Max}}$ is consistent with the data.
There is no reason why the other predicted curves would not be seen in the experiment and hence the breakdown scenario
is inconsistent with the data.  In particular, although it could be argued that the exact value of $B/B_0$ is uncertain
and hence there is  uncertainty in the relative size of the BD2 and BD1/3 orbits, there is no good reason why the
breakdown orbits at the top of the zone ($\rm{BD2_{max}}$) should be seen and not that at the center of the zone
($\rm{BD2_{min}}$). The breakdown gap is not strongly $k_z$ dependent.

\begin{figure}[ptb]
\center
\includegraphics*[width=8cm]{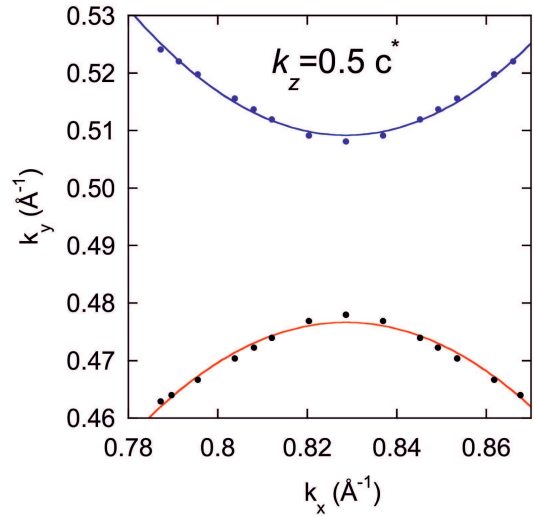}\caption{Fermi surface of the electron
sheets in LiFeAs close to the point where magnetic breakdown is likely to
occur (at $k_{z}=0.5c^{*}$). The solid lines are fits to the quadratic
expression given in the text.}%
\label{bandfit}%
\end{figure}

Finally, we address estimates of the size of the breakdown field. In Ref.\ \cite{borisenkocomment} it is suggested that
$B_{0}$ is very small. Although even that does not resolve the problem of reconciling the dHvA and ARPES data, we would
like to comment further on this. $B_{0}$ can be estimated from the band-structure data using the following formula due
to Chambers \cite{chambers}
\[
B_{0}=\frac{\pi\hbar}{2e}\left(  \frac{k_{g}^{3}}{a+b}\right)  ^{1/2}%
\]
where $k_{g}$ is the gap in $k$-space, and $a$ and $b$ are found from fits to the band dispersions close to the gap,
$\epsilon_{1}=\epsilon_{F}+\hbar v_{F_{1}}(k_{x}+\frac{1}{2}k_{g}+\frac{1}{2}ak_{y}^{2})$ and $\epsilon
_{2}=\epsilon_{F}+\hbar v_{F_{2}}(k_{x}-\frac{1}{2}k_{g}-\frac{1}{2}bk_{y}%
^{2})$. By fitting our band-structure calculations (see Fig.\ \ref{bandfit}) to
this expression we get (at $k_z=0.5c^*$) $k_{g}=0.030$\thinspace\AA $^{-1}$, $a=9.3$%
\thinspace\AA \ and $b=8.6$\thinspace\AA . This gives $B_{0}=125$\thinspace T and hence a damping factor of
approximately $\exp(-500/B)$ for the elliptic orbit (at $k_z=0$ we find $B_0=156$\thinspace T). Clearly, this is much
larger than our experimental fields explaining why we do not observe the magnetic breakdown orbits. The measured
damping factor for the $\varepsilon$ orbit in our sample \cite{putzke11} is approximately $\exp(-160/B)$. This is
consistent with impurity scattering alone (mean-free-path=400\thinspace\AA), leaving no room for any significant
contribution from magnetic breakdown.

It is well known that correlation effects (which are relatively strong in LiFeAs where $m^*/m_b \simeq 5$ for some
orbits) enhance SO and hence our estimated breakdown field is likely too \textit{low}. In the 122 compound it was found
that the DFT calculations underestimate the magnetocrystalline anisotropy (spin gap) by a factor of two \cite{spingap2}
to three \cite{spingap} (meaning the SO coupling is underestimated).  Also our analysis of LiFeP, where we have a
nearly complete experimental picture, indicates that our calculations underestimate SO in the hole bands, and by
implication, in the electron bands. In fact it is consistent with the fact that the ultra-small hole orbit 1 is not
observed in LiFeAs (as opposed to LiFeP), if SO is stronger in LiFeAs.

The ARPES spectra presented in the Comment \cite{borisenkocomment} do not have sufficient resolution to determine this
gap.  Different ARPES curves from the two electron bands are just about discernable at their maximum separation
($124^\circ$) but as the SO gap in our calculations is $\sim$ 7.5 times smaller than this it would be impossible to see
with the resolution shown in the figure.  In addition, `matrix' elements effects mean that the second electron band is
not observed for any angle less than 90$^\circ$.  It is unclear how the `matrix' elements evolve with angle so the SO
gap may be unobservable even if there were sufficient resolution.

To conclude, we would be very happy if our bulk dHvA were consistent with ARPES measurements, and in fact we had been
looking for such an interpretation. Unfortunately, this does not appear to be borne out by the data.

\end{document}